\documentclass{mnras}
\usepackage{graphicx}

\def\pg{{PG1211+143}}
\def\ngc{{NGC4051}}

\def\le{{L_{\rm Edd}}}
\def\msun{{\rm M_{\odot}}}

\def\me{{\dot M_{\rm Edd}}}

\def\xmm{{\it XMM-Newton}}
\def\chandra{{\it Chandra}}
\def\suzaku{{\it Suzaku}}
\def\et{{et al.\ }}
\def\swift{{\it Swift}}

\newcommand{\ls}{\mathrel{\hbox{\rlap{\hbox{\lower4pt\hbox{$\sim$}}}\hbox{$<$}}}}
\newcommand{\gs}{\mathrel{\hbox{\rlap{\hbox{\lower4pt\hbox{$\sim$}}}\hbox{$>$}}}}

\def\Msun{\hbox{$\rm ~M_{\odot}$}}

\def\H0{{\rm ~km~s^{-1}~Mpc^{-1}}}

\def\msun{M_{\rm \odot}}

\def\et{{et al.}}

\title[An ultra-fast wind in \pg]
        {Observing the launch of an Eddington wind in the luminous Seyfert galaxy \pg}
\author[Ken Pounds \et]
        {Ken Pounds and Kim Page\\
Department of Physics and Astronomy, University of Leicester, Leicester, LE1 7RH, UK \\}

\date{Accepted ; Submitted }
\pagerange{\pageref{firstpage}--\pageref{lastpage}}
\pubyear{2005}
\begin{document}
\maketitle
\label{firstpage}

\begin{abstract}

  The luminous narrow line Seyfert galaxy \pg\ was the first non-BAL AGN to reveal a powerful ionized wind, based on early
  observations with ESA's \xmm\ X-ray Observatory. Subsequent observations, mainly with \xmm\ and the Japanese \suzaku\
  Observatory, found such winds to be a common feature of luminous AGN. Typical outflow velocities of v $\sim 0.1$c and flow
  momenta mv $\sim\ L_{Edd} /c$ are consistent with winds being launched by continuum driving from a disc when the local mass
  accretion rate is super-Eddington. Here we report the launch of a new, ultra-fast outflow component in \pg\, near the end of
  a 5-week \xmm\ observing campaign, and discuss its origin in an ultra-fast {\it inflow} detected some 3 weeks earlier.
  We note that the inflow lasted for at least 3 days and delivered at least 10 Earth mass of fresh material into the innermost region of
  the source. While this mass by itself is insufficient to cause a complete inner disc restructuring - a prediction supported by lack
  of change in simultaneous UV fluxes - we suggest that a ring of matter 
  at $R\sim 20 R_{\rm g}$, located via its gravitational redshift (Pounds and Page 2024),
  was subsequently accreted, leading to the launch of a new outflow at a velocity of v $\sim 0.27$c
\end {abstract}

\begin{keywords}
accretion, accretion discs -- galaxies: active -- quasars: general -- galaxies: individual: PG1211+143 -- galaxies:Seyfert -- X-rays:galaxies 
\end{keywords}{

\section{Introduction}
X-ray spectra from an \xmm\ observation of the narrow-line Seyfert galaxy \pg\ in 2001 provided the first detection in a non-BAL
AGN of strongly blue-shifted absorption lines of highly ionized gas, corresponding to an outflow velocity of
0.15$\pm$0.01c (Pounds \et\ 2003, Pounds and Page 2006). Further observations over several years with \xmm, \chandra\ and
\suzaku\ showed the high velocity outflow to be persistent but of variable opacity (eg Reeves \et\ 2008). Evidence that
the extended outflow in \pg\ was both massive and energetic - with potential importance for galaxy feedback - came from the
detection of PCygni and other broad emission features obtained by combining the 2001, 2004 and 2007 \xmm\ EPIC spectra (Pounds 
and Reeves 2007, 2009). 

Examination of archival data from \xmm\ and \suzaku\ has since shown similar ultra-fast, highly-ionized outflows (UFOs) to be
relatively common in nearby, luminous AGN (Tombesi 2010, 2011; Gofford 2013). The frequency of these detections confirms a
substantial covering factor and hence significant mass and kinetic energy in such winds. Indeed, their integrated mechanical
energy may be substantially greater than required to disrupt the bulge gas in the host galaxy, suggesting some winds are intermittent,
or that much of the energy in a persistent wind must be lost before reaching the star forming region, perhaps by colliding with pre-ejecta,
as seen in an \xmm\ observation of the low mass Seyfert \ngc\ (Pounds and Vaughan 2011, Pounds and King 2013).  

In order to further explore the velocity structure and evolution of the fast wind in \pg\ an extended \xmm\ observation was
carried out during 7 spacecraft orbits  over the period 2014 June 2 to 2014 July 9. Effective on-target exposures for individual
orbits ranged from $\sim$50 to $\sim$100 ks, with a total duration of $\sim$650 ks. Full details of the \xmm\ observing log are
given in Lobban \et\ (2016), reporting the results of a detailed timing analysis. 

The {\it Neil Gehrels Swift Observatory} (\swift; Gehrels et al. 2004) also observed PG 1211+143 with 43 snapshots, as part of a Target of Opportunity (ToO) programme linked to the
\xmm\ campaign. The \swift\ observations cover the period 2014 June 4 to 2014 August 4 and have
typical durations of $\sim$ 1.5 ks. We have extracted X-ray light curves from the X-ray Telescope (XRT; Burrows et al. 2005) and
UV light curves from the Ultra-Violet/Optical Telescope (UVOT; Roming et al. 2005),  the latter using the u, uvw1 and uvw2
filters. Full details of the \swift\ observations are also presented in Lobban et al. (2016).

Figure 1 reproduces the orbital-mean X-ray fluxes from the \xmm\ pn camera (Strueder \et 2001), together with the first 17 snapshots from
the \swift\ XRT. Both data sets show a deep minimum flux near \xmm\ orbit 2659 (day 16), when the ultra fast inflow was
detected. The \swift\ light curve is particularly interesting, with its softer X-ray bandwidth responding to both column density and
ionization changes in the line-of-sight flow, suggesting the inflow seen in day 16 actually began some 3 days earlier, followed by a strong
increase in X-ray emission, to a peak in orbit 2664 some 8 days later. The high Xray flux is then maintained for at least 4 further days,
to orbit 2666, but has fallen substantially by the final observation (orbit 2670).

\begin{figure}                                                          
\centering                                                               i
\includegraphics[width=8cm, angle=0]{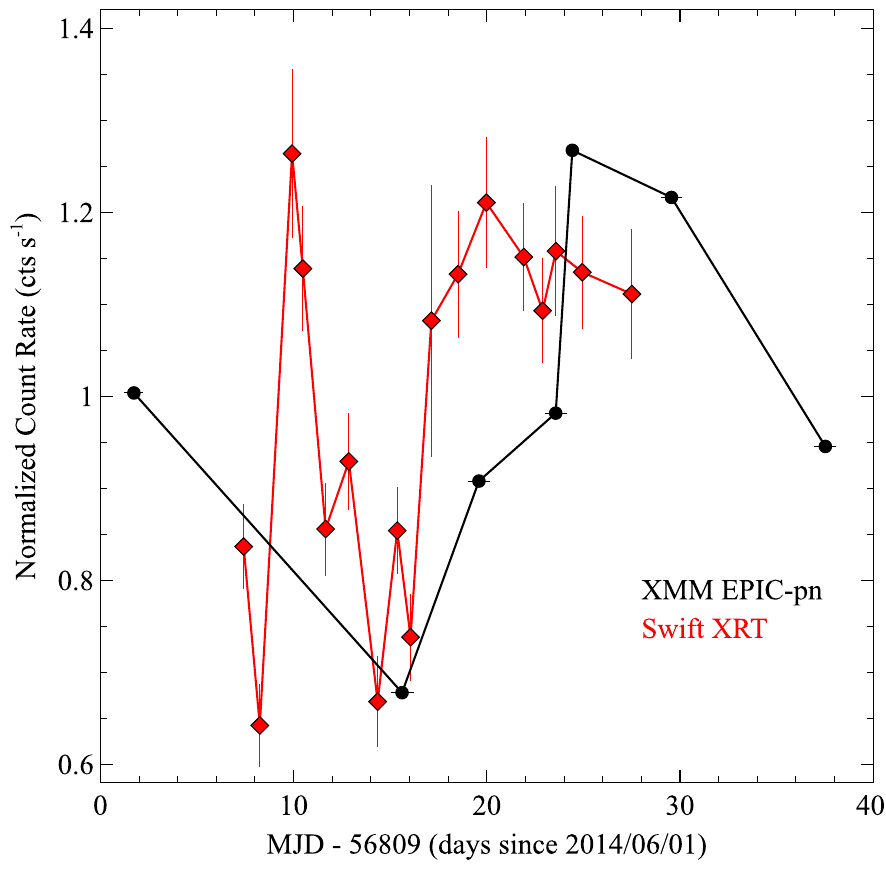}                              
\caption                                                                
{X-ray count rates from the pn camera for each of the 7 observations in the 2014 \xmm\ campaign, together with 17
 overlapping observations from the \swift\ satellite. \xmm\ data points correspond to spacecraft orbits 2652, 2659, 2661, 2663, 2664, 2666
 and 2670.
 Both data sets show a  deep minimum flux near orbit 2659 (day16), followed by an increase to a peak near orbit 2664 (day24). The earlier
 recovery of X-ray flux in the \swift\ data is explained by the softer XRT energy band, compared with that of the \xmm\ pn camera, as the
 rising continuum increases the ionization state of the inflowing matter.}      
 \end{figure}  

Figure 2 shows X-ray spectra from the \xmm\ pn camera over the same interval, with orbits 2659 (black), 2661 (red), 2663 (green) and
2664 (blue), plotted as a ratio to that of orbit 2652. The broad spectral band highlights the strong soft X-ray absorption associated
with the transient accretion event in orbit 2659, which then falls over several days, with the X-ray emission increasing to orbit 2664
as some of the new matter is
accreted.

\begin{figure}                                                          
\centering                                                              
\includegraphics[width=10cm, angle=0]{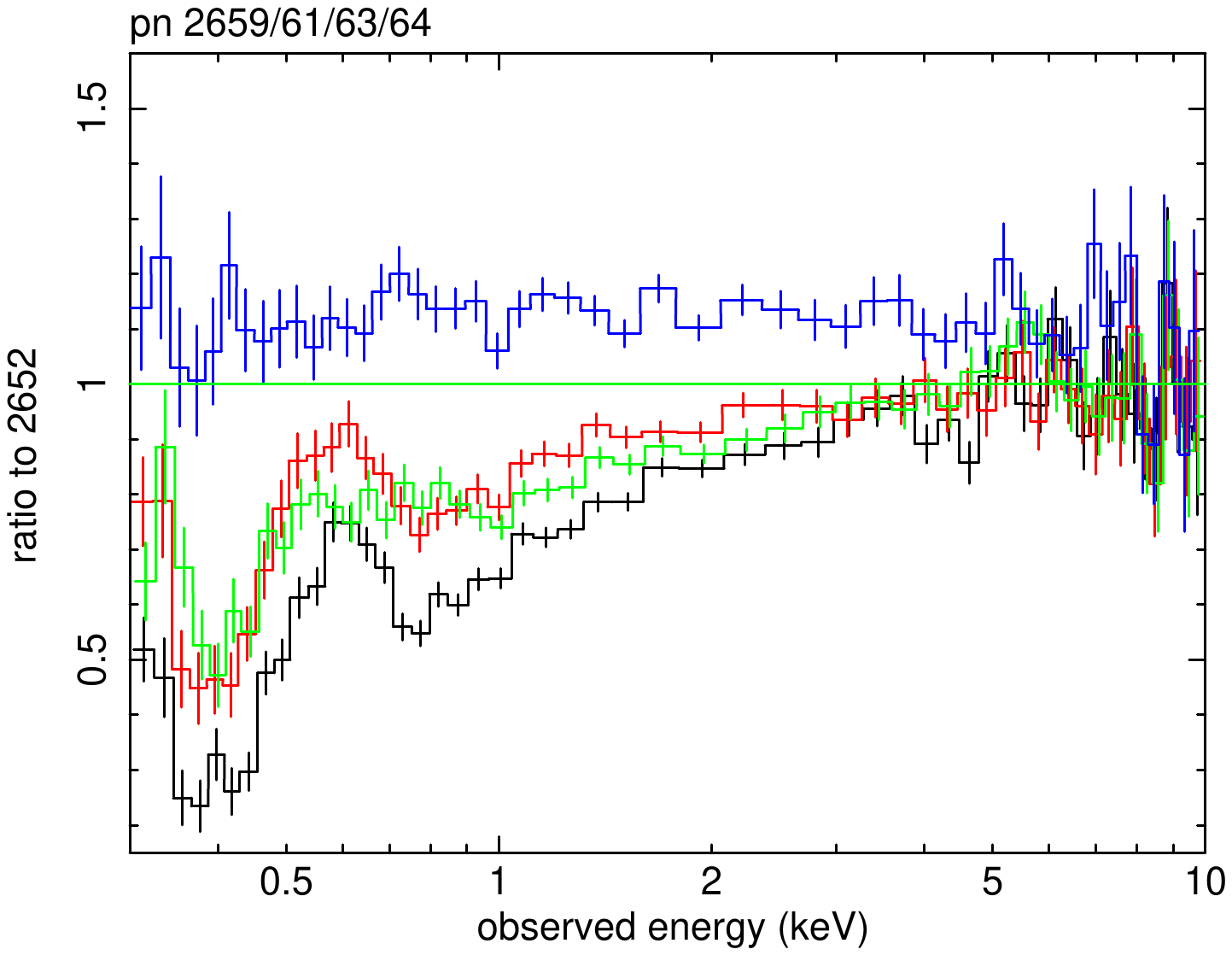}
\caption                                                                
{Broad-band spectra from the \xmm\ pn camera for orbits 2659(black), 2661(red), 2663(green) and 2664(blue) plotted as a ratio to that of
orbit 2652, illustrating strong soft X-ray absorption during the transient line-of sight inflow on day 16, then falling while the
X-ray emission increases - as additional matter is accreted - to a new peak in orbit 2664.}      
\end{figure}  

Published analysis of the 2014 \xmm\ observation of \pg\ has focussed on stacked X-ray spectra, where the  
high quality data have revealed a more complex velocity structure, with primary (high column density) outflow components at 
v$\sim$0.066c and v$\sim$0.129c detected in both pn and RGS spectra (Pounds \et\ 2016a, 2016b; hereafter P16a and P16b).

Notably, neither outflow velocity in 2014 was consistent with that of v$\sim$0.15c in the initial \xmm\ observation in 2001, a not uncommon
finding where repeated observations of UFOs reported in the archival searches showed differing velocities weeks apart, implying that at
least some high velocity AGN winds may be relatively short-lived. An initial examination of individual orbits during the 5-weeks
\xmm\ observation of \pg\ in 2014 found the clearest spectral variability in the soft X-ray band, sensitive to both column density
and ionization state changes, with a detailed inter-orbit analysis of the RGS soft X-ray spectra confirming variability on timescales of
days, and the strongest outflow at v$\sim$0.06c clearly resolved into distinct ionization (density) components (Reeves \et\ 2018).

The present paper derives from an orbit-by-orbit study of the harder X-ray spectra from the EPIC pn (Strueder \et\ 2001) and MOS
(Turner \et\ 2001) cameras. One remarkable outcome previously reported (Pounds \et\ 2018) was the detection of a transient ultra-fast {\it
 inflow} velocity, with v$\sim$0.3c, during the second \xmm\ orbit in 2014. We report here the launch of a new high velocity {\it outflow} of v$\sim$0.27c
several weeks later, and suggest the two events are causally linked, a supposition supported by the detection in stacked 2014
spectra of a ring of matter very close to the supermassive black hole (SMBH; Pounds and Page 2024).

We assume a redshift of $z=0.0809$ (Marziani \et\ 1996), with a black hole mass of $4\times 10^{7}$\Msun\ (Kaspi \et\ 2000) indicating the
historical mean luminosity of \pg\ is close to Eddington. Spectral modelling is based on the XSPEC package (Arnaud 1996) and includes
absorption due to the line-of-sight Galactic column N$_{H}$ $\sim 3\times10^{20}$ (Murphy \et\ 1996). To allow comparison with previous
analyses of the 2014 spectra, in particular P16a, we again use photoionized absorption (grid25) and emission (grid 22) arrays of pre-computed spectra,
based on the XSTAR code of Kallman \et\ (1996).

\section{Launch of a third high velocity outflow late in the 2014 campaign}                                                   

As noted above, the unusually high statistical quality of the extended 2014 observation of \pg\ was important in resolving complex
absorption structure in the Fe K spectrum, identified in P16a with absorption line series in FeXXV and XXVI corresponding to line-of-sight
outflow velocities of v$\sim$0.067c and v$\sim$0.129c, and subsequently confirmed in the soft X-ray spectrum reported in P16b. Table 1 lists
the parameters of the 2 primary (high column) outflows from the mean 2014 pn spectral fit.

\begin{table*}
\centering
\caption{Parameters of the ultra-fast outflow obtained from a 2--10 keV spectral fit to the stacked 2014 pn data, with 2
photoionized absorbers defined by ionization parameter $\xi$ (erg cm s$^{-1}$), column density $N_{\rm H}$ (cm$^{-2}$) and
outflow velocity ($v/c$). Absorber 3 relates to the new outflow reported in the present paper. Extracted luminosities (erg s$^{-1}$) for each
photoionized component are over the 2--10 keV observed energy range.}
\begin{tabular}{@{}lccccc@{}}
\hline
comp & log$\xi$ & $N_{\rm H}$($10^{23}$)  & $v/c$ & $L_{\rm abs}$ & $\Delta \chi^{2}$ \\
\hline
abs 1 & 3.5$\pm$0.1 & 2.6$\pm$0.2  & 0.066$\pm$0.003 & 11$\times10^{41}$ & 52/3 \\
abs 2 & 3.9$\pm$0.6 & 1.5$\pm$0.2  & 0.129$\pm$0.006 & 6$\times10^{41}$ & 18/3 \\
abs 3 & 2.9$\pm$0.2 & 1.0$\pm$0.3  & 0.27$\pm$0.01 & 4$\times10^{41}$  & 17/3 \\
\hline
\end{tabular}
\end{table*}

Examination of pn data for individual orbits shows those 2 outflow velocities remained essentially as in Table 1, indicating both
were in `coasting' phase, a common property expected of winds launched by continuum radiation (King and Pounds 2003, 2015).

However, in the final 2014 observation (orbit 2670) several additional absorption lines appeared, notably at $\sim$8.2 kev and $\sim$8.6 keV
and at 9-10 keV, indicating absorption specific to orbit 2670.  To quantify the new absorption component, the spectral model from the
stacked 2014 data (P16a) was applied to the pn data from orbit 2670, with only the normalisation (amplitude) of X-ray continuum and
emission components free to change.  Figure 3 (upper panel) shows the additional absorption in the ratio of orbit 2670 data to the mean 2014
spectrum.

Adding a 3rd photoionized absorber to the spectral model - with other parameters frozen - confirms a new, high column, outflow, at an observed blueshift of 0.185$\pm$0.005,
with column density N$_{H}$ = (9.7$\pm$2.8)$\times10^{22}$cm$^{-2}$ and ionisation parameter log$\xi$ $\sim$2.9.  The corresponding outflow
velocity in the AGN rest frame is v$\sim$0.27$\pm$0.01c.  Adding the additional absorber (Figure 3, lower panel) recovers an excellent
spectral fit to the orbit 2670 data, with $\Delta \chi^{2}$ of 17/3 and a null probability of 4$\times10^{-3}$.

\begin{figure}                                                          
\centering                                                              
\includegraphics[width=9cm, angle=0]{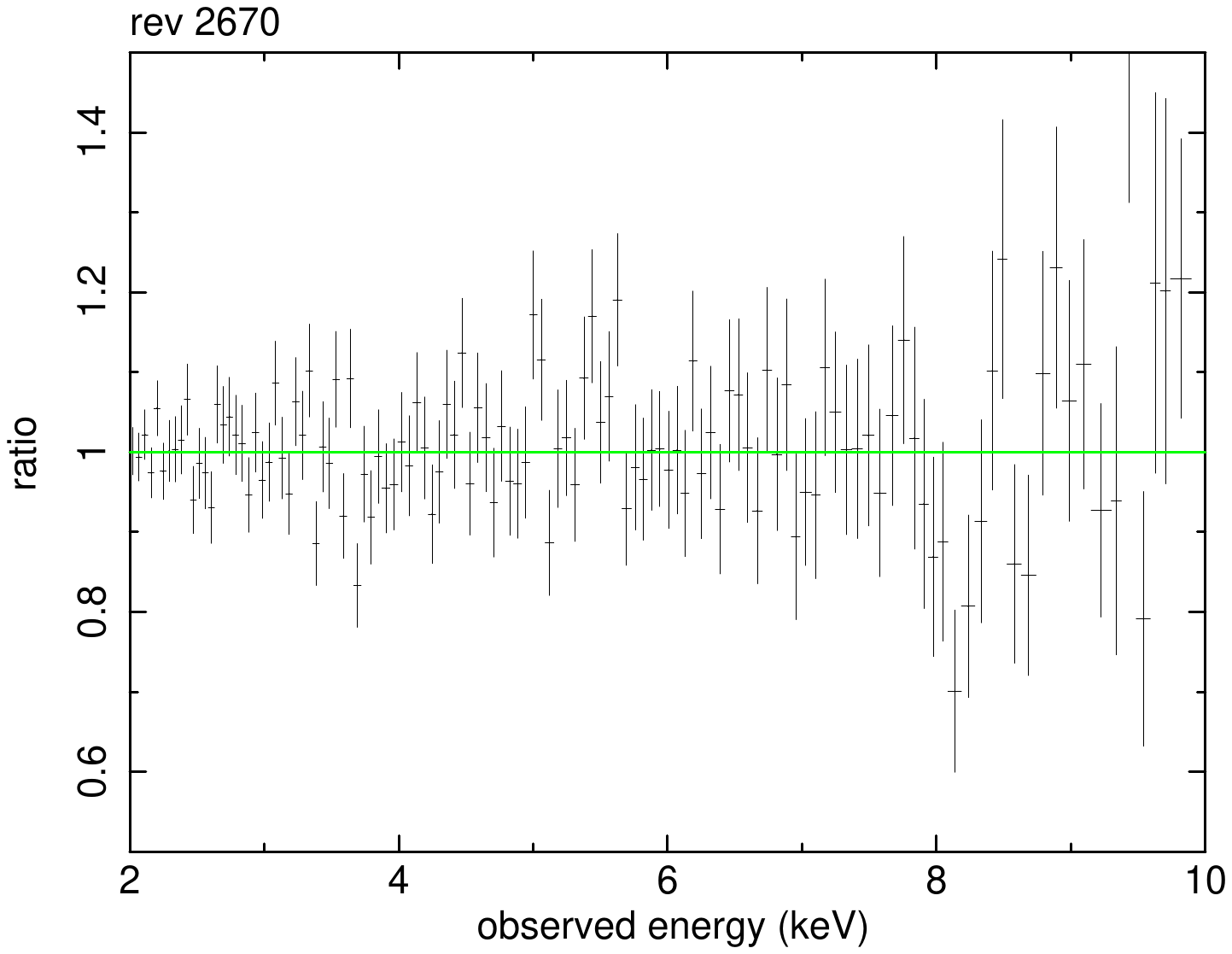}                                           
\includegraphics[width=9cm, angle=0]{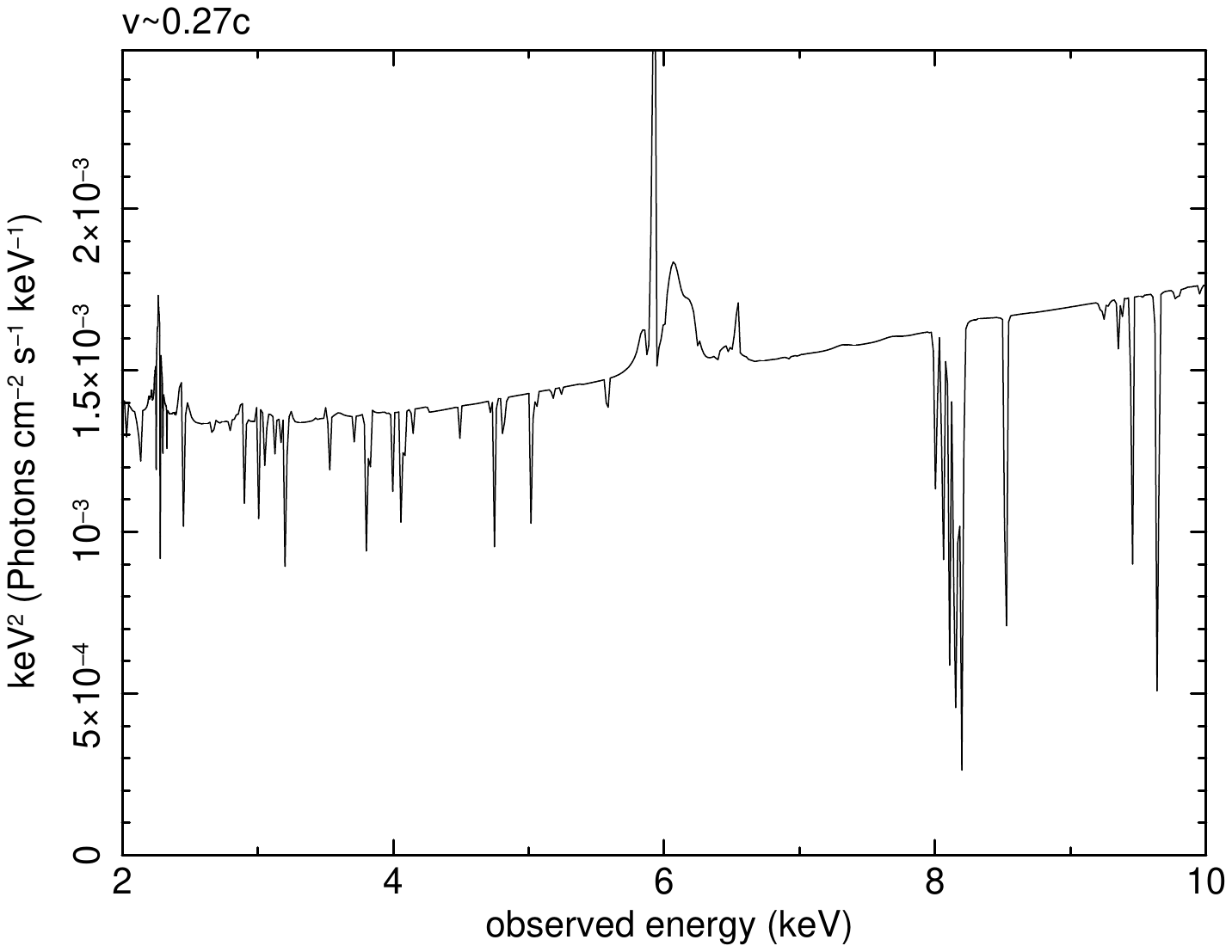}                                           
\caption  {(top) Absorption features specific to rev2670 are shown by fitting the rev2670 pn data with the 2014 mean spectrum, allowing
  only the normalisation of continuum and photoionized emission to vary. New absorption lines are seen near 8.2 and 8.6 keV, with
  additional spectral structure between 9 and 10 keV.
  (lower) Absorption profile of a 3rd primary outflow component with column density N$_{H}$ $\sim$ $10^{23}$, ionization parameter
  log$\xi \sim$ 2.9 and observed blueshift $\sim$ 0.185}                              
\end{figure}

Comparison of the upper and lower panels of Figure 3 identifies the resonance lines of He-like FeXXV and H-like FeXXVI blue-shifted to
8.2 and 8.6 keV, respectively, and the corresponding $\beta$ transitions. The red wing in the $\sim$8.2 keV line is explained by inner-shell
absorption on the low energy wing of the Fe XXV resonance line, a feature often seen in AGN spectra.

Independent support for the new high velocity outflow is provided by a similar examination of the RGS data from orbit 2670, where a
comparison with the mean 2014 spectrum (P16b) again reveals additional soft X-ray absorption specific to orbit 2670, at a redshift of
0.183$\pm$0.003, with column density N$_{H}$ = (9.5$\pm$3.0)$\times10^{21}$cm$^{-2}$ and ionisation parameter log$\xi\sim$0.8.  The additional
absorber provides a good spectral fit to the orbit 2670 RGS data, with $\Delta \chi^{2}$ of 12/3 and null probability
of 7$\times10^{-3}$.

\begin{figure}                                                          
\centering                                                              
\includegraphics[width=10cm, angle=0]{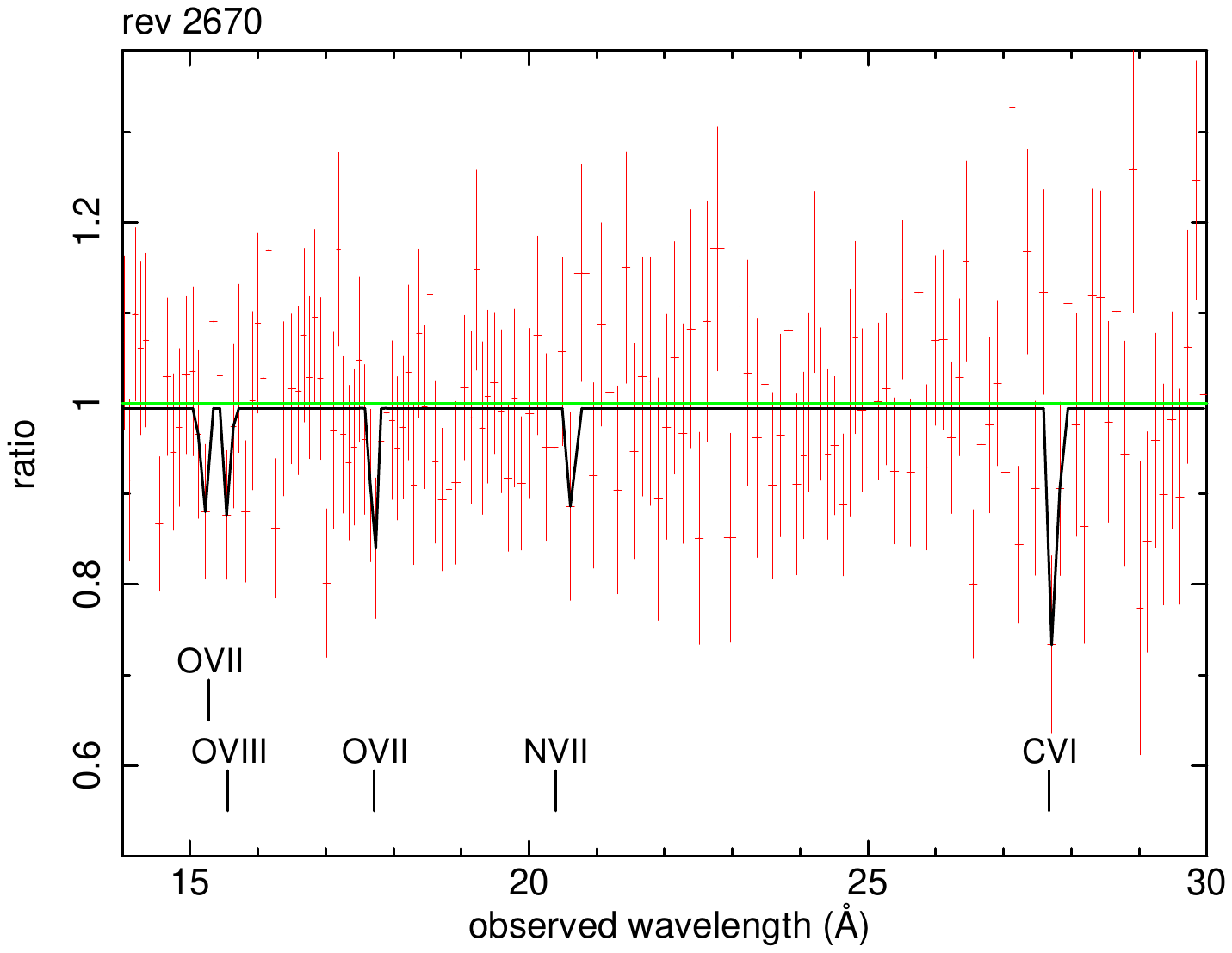}                                           
\centering                                                                                                         
\caption  {Soft X-ray absorption specific to orbit 2670 is identified with blue-shifted resonance lines of OVII, OVIII, NVII
 and CVI, all consistent with an outflow velocity in the AGN rest frame of v$\sim$0.265}
\end{figure}

Figure 4 shows an array of soft X-ray absorption lines specific to orbit 2670, identified with resonance transitions in OVII, OVIII, NVII
and CVI, with a common blueshift $\sim$ 0.185$\pm$0.006, corresponding to an outflow velocity v$\sim$0.265$\pm$0.005c. The lower ionization
parameter and column density (compared with the pn data) indicate higher density matter probably embedded in the primary (high column)
highly-ionized flow.

\section{Discussion}

Repeated \xmm\ observations have demonstrated the highly variable nature of the ultra-fast outflow in \pg. The initial discovery of a
strong outflow at 0.15c in a 2001 \xmm\ observation was followed by weaker detections in 2004 and 2007, and then the long 2014 campaign
reporting a more complex velocity structure.  Assuming the highly ionized winds are driven by momentum exchange with a super-Eddington
radiation flux (King and Pounds 2003), an explanation supported in the quantitative agreement of data and theory (King and Pounds 2015), the
pattern of wind variability will also have direct bearing on the nature of variable accretion in AGN.
 
Nixon \et\ (2012) have shown how an accreting `cloud' approaching the disc at an oblique angle to the black hole spin plane could cause the
inner disk to warp and break off, with subsequent collisions between rings of matter precessing at different rates, leading to loss
of angular momentum and direct infall, and potentially creating local disc regions of super-critical mass accretion. The ultrafast infall
detected during the 2014 \xmm\ observations of \pg\ was discussed in that context in Pounds \et (2018; P18).

Observing the launch of a new primary (high column) ultrafast outflow component, some 20 days later, is a striking illustration of
continuum-driving for a high velocity wind, as first described in the classic accretion disc paper of Shakura and Sunyaev (1973), but -
interestingly - not mentioned in the original scientific case for \xmm\ (Bleeker \et\ 1982).

In that context, the near coincidence of a 0.3c infall velocity (with free-fall location at $\sim$20$R_{\rm g}$) detected in rev 2659, and
the wind launch seen 2-3 weeks later suggests a physical link.  That
speculationn is strengthened by the recent report of a ring of matter orbiting the SMBH, most likely populated by the transient
inflow, at a radius (determined by its gravitational redshift) of $\sim$22$R_{\rm g}$ (Pounds and Page 2024).

We suggest the increase in X-ray emission between those two events (Fig.2) then reflects the release of energy as the new matter reaches the
inner disk region. 

To examine that idea quantitatively, we recall the transient inflow mass reported in P18 was $\sim$3.3$\times10^{26}$g,
accumulated over 3000s, for a mean rate of $10^{23}$g $s^{-1}$.  As that measure is only for matter in line of sight and sufficiently cool
to be detected, it is likely to be a substantial underestimate. Furthermore, Fig. 1 suggests the infall may have begun some 3 days
earlier, increasing the observed mass dump by a further factor.

As the Eddington accretion rate for the source is

\begin{equation}
    \me = \frac{\le}{0.1 c^2} = 0.8 \; \msun {\rm yr}^{-1} 
\end{equation}

or $\sim$~6 $\times$ 10$^{25}$g~s$^{-1}$, it appears unlikely that the ultra-fast inflow would have significantly affected the {\em bulk} of
the accretion flow - the optically thick UV-emitting region. That prediction is now supported by observation, with the UV fluxes from
\swift\ over the same period showing no substantial variability for several days before and after the ultra fast inflow seen on day 16
(Fig.5).

In contrast, the X-ray data from \swift\ and \xmm\ are highly variable (Figs. 1 and 2), with enhanced soft X-ray absorption between days 12
and 15 (coincident with the fast inflow), followed by an increase in the broad band X-ray emission peaking near day 24 (rev2664).  Unfolded
pn spectra (Fig.6) compare the power law componens (identified as emission from the inner disc and corona) for rev2652, before the fast
transient inflow; rev2664, some 10 days later; and on the final observation, rev2670, coincident with the new wind launch.
The respective x-ray (2-10 keV) luminosities, of 6.22, 7.01 and 5.58 ($\times$ $10^{43}$erg $s^{-1}$) indicate the post-inflow strengthening of the
coronal emission, and a subsequent fall coincident with the launch of the new outflow.

\subsection{Coronal disruption}

In an earlier post on astro-ph (arXiv: 2310.18105) we suggested the additional material and extra energy introduced by the transient infall reported in 2018 may
have caused a significant rearrangement in the corona, where the top layers of the disc produce most of the high energy disc emission, thereby
leading to the launch of a new wind late in the 2014 campaign.
The inflow is certainly massive and powerful enough to have affected the corona of the accretion disc in \pg. Disc coronae must be
Compton optically thin to explain typical AGN spectra (e.g., Poutanen \& Svensson 1996). This implies that the surface column depth in
the corona is $\Sigma_c \leq \tau/\kappa_{\rm es} \sim 2$, where $\tau_c$ and $\kappa_{\rm es}$ are the corona optical depth and electron
scattering opacity coefficient, respectively. The effect of absorption in a line-of-sight inflow on the \swift\ XRT data
(Fig. 2) suggests a substantial flow lasted for at least $\Delta t$ = 3~days, adding $ \sim \dot M \Delta t/(\pi R^2)/2 \sim $ a few
g/cm$^2$, per side of the disc, exceeding the expected amount of the material in the corona. Furthermore, the specific energy injected by
the inflow into the coronal region of the disc, $v_r^2/2 = 0.05 c^2$ is at least an order of magnitude higher than that of the X-ray emitting
coronal gas, $\sim 3 k_{\rm B} T_c/m_p$, where $k_{\rm B}$ is the Boltzmann's constant, $T_{\rm c} \ga 10^9$~K is the corona
temperature, and $m_p$ is proton mass.

\subsection{Accretion specific to the innermost disc}
While coronal disruption remains a possible origin for the new outflow, a more conventional explanation is now suggested by the detection
of a substantial ring of matter accumulated during the transient inflow, where the assumption of a gravitational redshift (of 0.123)
locates the ring at $R\sim 20R_{\rm G}$ (Pounds and Page 2024).  
We now propose that the 0.27c wind was launched as a direct result of accretion from that inner ring of matter - at
a locally super-Eddington rate - with excess matter then being ejected by radiation pressure (King and Pounds 2003).
The post-inflow increase in X-ray emission to orbit 2664 - and subsequent fall in orbit 2670 - (Figs 2 and 6) provide another indicator
of the changes in the inner disc.

\begin{figure}                                             - 
\centering                   
\includegraphics[width=10cm, angle=0]{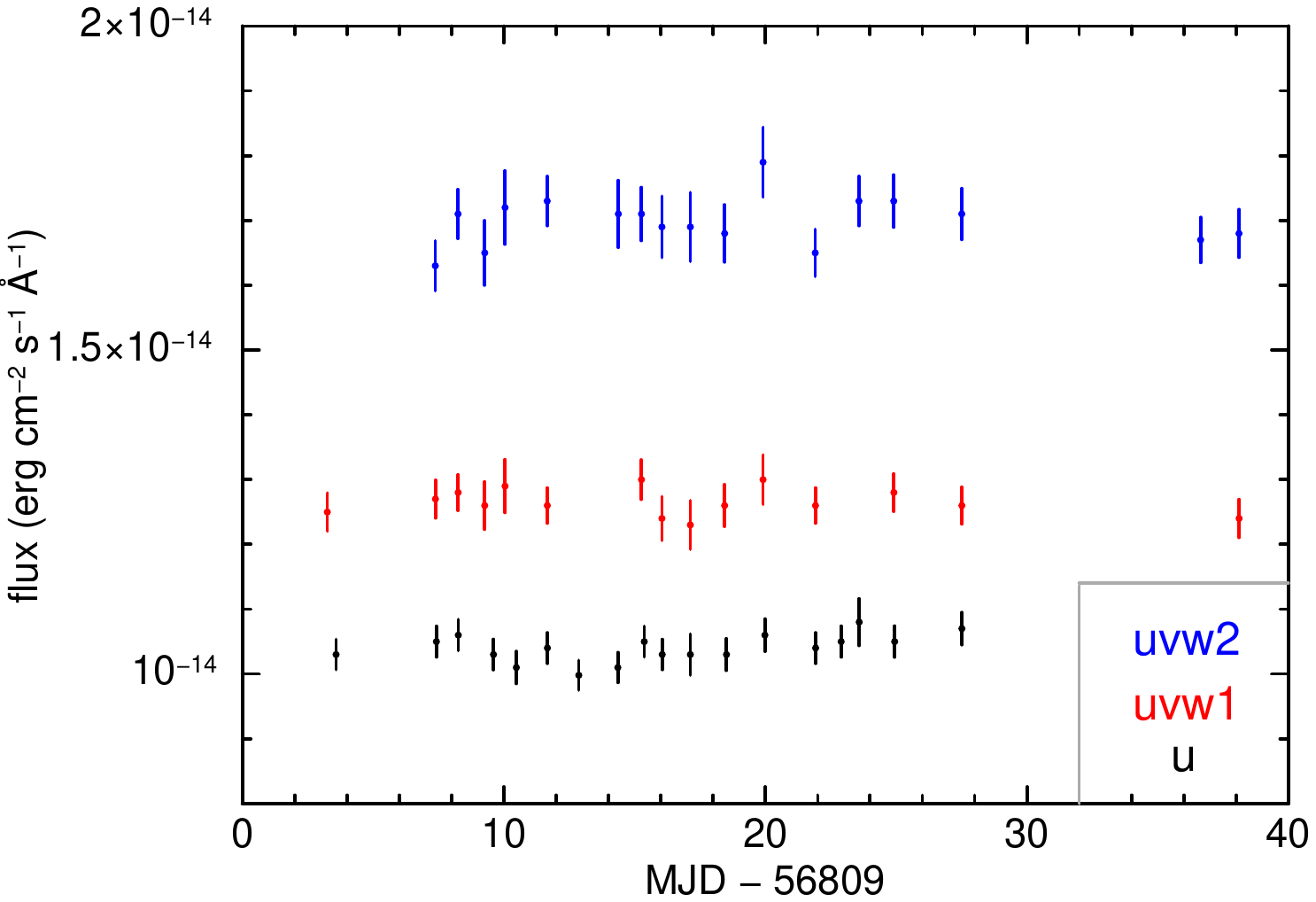}
\caption{UV observations from the \swift\ satellite showing no significant flux variability for several days before and after the ultra
 fast inflow observed on day 16, in contrast to the X-ray spectra over the same period. Black, red and blue data points show u, uvw1 and uvw2
 fluxes.}      
\end{figure}

\begin{figure}                                                          
\centering                                                              
\includegraphics[width=10cm, angle=0]{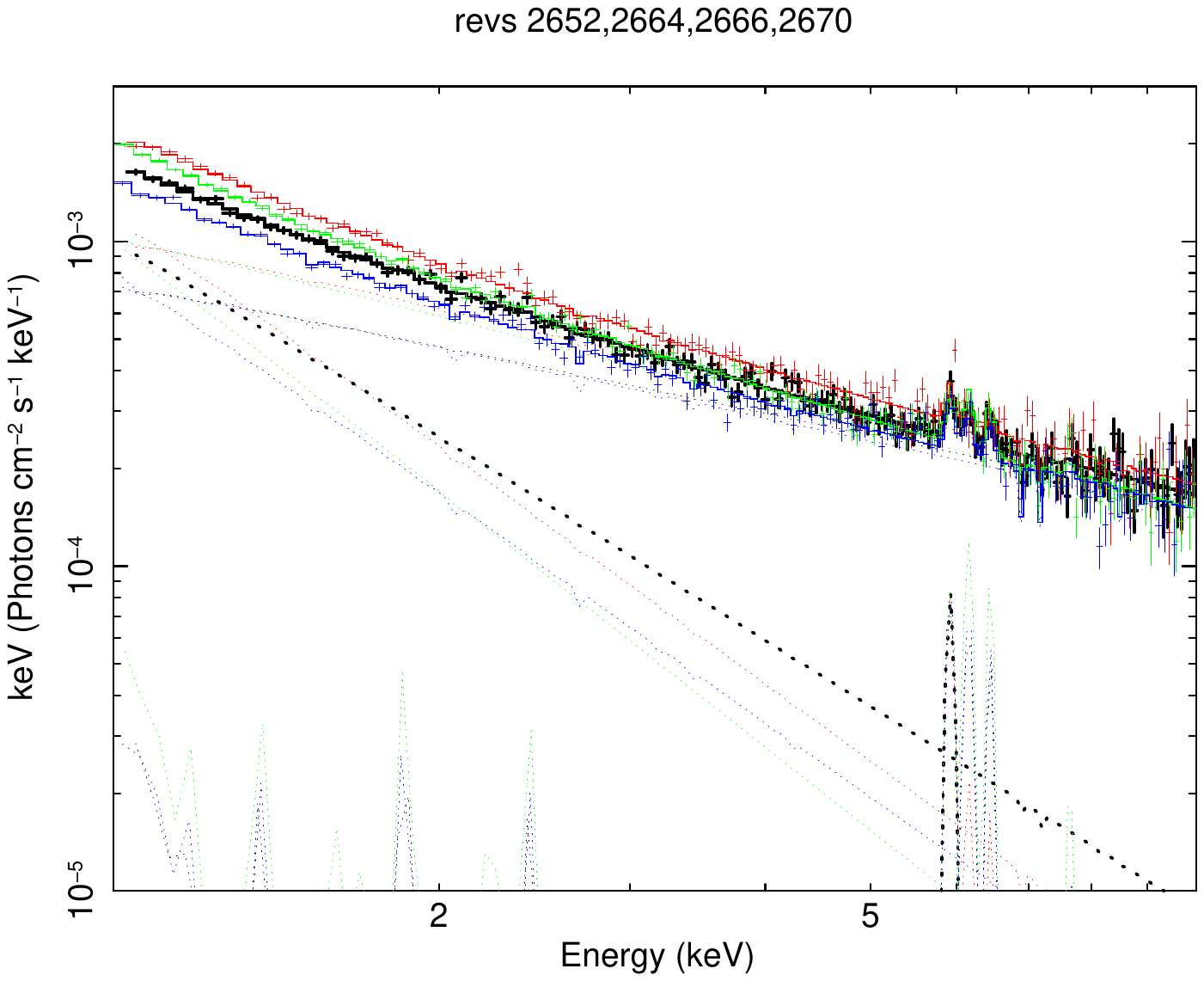}                                           
\centering                                                                                                         
\caption{Unfolded pn X-ray spectra: from orbit 2652 (black),some 10 days before the fast transient inflow; orbits 2663 (green) and 2664 (red),
 as the post-inflow X-ray emission increased and - finally - orbit 2670 (blue), coincident with the launch of a new wind.}
\end{figure}

\section{Conclusions}

Our analysis indicates that the amount of material delivered into the innermost tens of $R_{\rm g}$ by the inflow is likely to be
insufficient to disrupt the optically thick disc in that region, a conclusion supported by the lack of significant variability in the UV
data from \swift\ (Fig.5).  However, increasing soft X-ray emission from \xmm\ orbits 2659 to 2664 (Fig.2 and Fig.6; in Fig. 6, the soft and hard power laws relate to the enhanced inner disc emission, and that upscattered by the hot coronal plasma) is consistent with 
additional mass and energy being injected into the inner disc following the transient inflow detected earlier in the 2014
\xmm\ campaign and are the direct cause of a new UFO
launched some three weeks later. 

\section{ Data availability }
The data underlying this article are available in the XMM archive at http://nxsa.esac.esa.int/nxsa-web . The \swift\ data are available from https://www.swift.ac.uk/swift\_live .

\section*{ Acknowledgements }
\xmm\ is a space science mission developed and operated by the European Space Agency. We acknowledge in particular the excellent
work of ESA staff in Madrid successfully planning and conducting the \xmm\ observations. We also thank the \swift\ PI for
approving, and NASA mission planners for scheduling, the additional observations used in this paper. \xmm\ data were reduced by our former
colleague Andrew Lobban and \swift\  data were
provided by the UK Science Data Centre at the University of Leicester, supported by the UK Science and Technology Facilities
Council. We are grateful to Sergei Nayakshin for insightful comments on an earlier version of this paper.

\end{document}